# A Quantum Computing Framework for Complex System Reliability Assessment

Shutang You
University of Tennessee, Knoxville, TN, USA
Email: syou3@utk.edu

*Abstract*— **This paper proposed a framework based on quantum computing for reliability assessment of complex systems. The "Quantum Twin" concept was also proposed. The framework can be used to accelerate the reliability assessment of large-scale complex systems, which could take much computation time for classical computers to achieve accurate results. Power system is used as an example of complex systems to illustrate the framework.**

## I. Introduction

Complex systems are systems that have many interacting components. Complex systems are becoming more common with the development of cyber-physical systems, Internet-of-things, and advanced manufacturing technologies. The risk and reliability assessment of complex systems is very intensive computationally due to the large amount of components and their unique failure mechanisms, as well as the complicated interaction between components.

The power grid is a practical example of complex engineered systems (Figure 1). The reliability of power grid is critical for almost all aspects of the modern society. Accurate assessment of power system reliability is important for development and economic analysis of control strategies. With the increase of renewable generation, other resources, environmental and weather uncertainties in the system, the real-time situational awareness and predictive assessment of power grid reliability is also becoming increasingly important.

Power grids are comprised of many components, including generators, transmission lines, transformers, reactive power devices, loads, etc., and each component may fail due to due to internal defects or environmental impacts. Large-scale power grid reliability assessment usually requires Monte Carlo simulation, which requires a large number of sampling and calculation, to obtain an approximate results of reliability assessment. The accuracy of Monte Carlo simulation mainly depends on the number and quality of Monte Carlo sampling. For a large-scale system, the required number of sampling is very large and it will take months or even years to capture low-probability but high-impact system states. Some recent studies utilized distributed computing and parallel computing to accelerate this calculation [1, 2], which showed significant improvements. However, as the uncertainty of the system increases and the requirement of real-time system reliability assessment emerges in the near future, power system reliability assessment may require an even faster computation technology.

In spite that quantum computing is still far from actual applications, quantum computing has made significantly progress in recent years [3]. Although any problems that quantum computers can solve can also be solved by classical computers theoretically, quantum computer requires much less computation time in some computation tasks compared with classical computers, which is often referred to "quantum supremacy". Quantum computing has been demonstrated a supremacy over classical computers in certain computation tasks in a lab environment [4]. It has also been used in improving machine learning performance [5]. Some previous work has explored its application in power systems [6], such as optimization [7] and forecasting [8]. So far, no study has looked at how quantum computing can improve the reliability assessment of complex systems like power grids. In fact, quantum computing can be a revolutionary tool for complex system reliability assessment.

This paper proposed a framework using quantum computing for reliability assessment of complex systems. Specifically, power system is used as an example of complex system to illustrate the framework. The "Quantum Twin" concept is proposed to explain the mechanism of using quantum bits and quantum logical gates to represent uncertainties of components states and interactions of components. Future work on using quantum computing for practical reliability analysis of large-scale complex systems is also discussed.

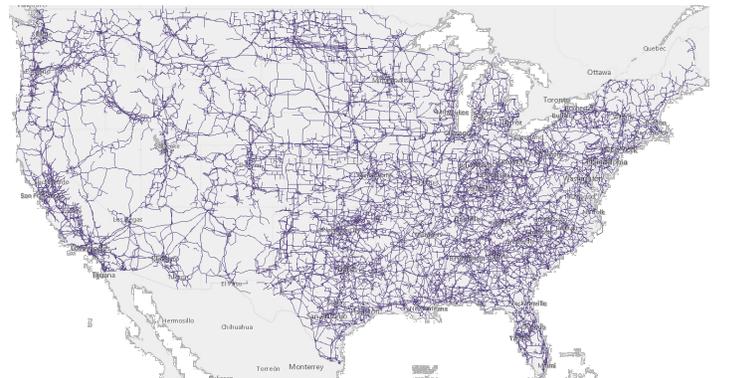

Figure 1. An example of complex systems: the U.S. continental interconnection power grid.

## II. Quantum Computing and Complex System Reliability Assessment — Example of Power Systems

### A. Background

*Qubit*: A qubit is the basic unit of quantum computing. It can be in a superposition of two states. Capable of denoting both the "on" and "off" of a component in a computation, a qubit is an ideal representation of the component states for complex system reliability assessment.

_Quantum Logic Gate_: A quantum logic gate is a basic quantum computation unit that takes several qubits as inputs. Unlike a classical logic gate based on classical circuit, a quantum logic gate is a reversible computation logic, which can intake dual states of components as inputs, accelerating the computation of complex system reliability.

_Quantum Computing_: Quantum computing uses qubits to represent component "on" and "off" states at the same time and uses quantum logic gates to accomplish analysis of systems that have many interacting components.

_Power System Reliability Assessment_: Power system reliability assessment is to measure the capability of power systems to serve customer demand. Its reliability depends on the availability of generation, transmission, and distribution resources. The reliability assessment is usually a statistical value considering the uncertainties of failures.

### B. Quantum Computing in Power System Reliability Assessment

Since the superposition of a qubit can represent two states at the same time, quantum computers can do many calculation considering different states of many components at the same time. Therefore, quantum computers are faster when doing multiple operations to search for the right answer. This feature is especially useful for the assessment of large-scale system reliability. If each qubit can represent both the "on" and "off" state of a component in power systems at the same time, then, for a quantum computer that has N qubits, it can represent a system that consider all combination of on/off states of N components. That is $2^N$ states of the system. This will significantly reduce the dimensionality disaster in computation for reliability evaluation.

Figure 2 shows a comparison between the classical reliability analysis and quantum computing based reliability analysis of power systems. In conventional reliability analysis, component states are represented by 0 and 1, which are then represented by the "switching on" and "off" states of electronics in the digital computation chip. The corresponding technology can be called "digital twin" in reliability assessment, in which digital bits are used to create a representation of the power system. In contrast, for the quantum computing based power system reliability assessment, the superposition of quantum bits enables representation of both in-service and out-of-service state possibilities in reliability computation. This technology can be called "Quantum Twin" in reliability assessment. A "Quantum Twin" is representation of a system, in which each component is represented by a qubit. The interaction or interrelationship between each component is represented by a quantum logic gate.

It should be noted that this framework also works for reliability assessment where time-domain simulation is needed. The process is similar to the case of conventional power flow based reliability assessment except for that quantum logic gates are formulated using algebra and differential equations for time-domain simulation.

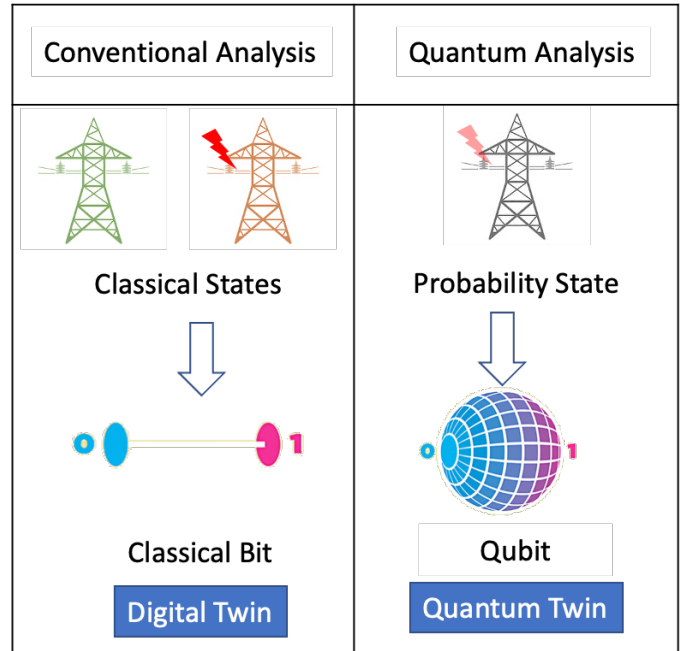

Figure 2. Quantum analysis of power system reliability assessment

### III. DISCUSSION ON FUTURE WORK

Because of its unique capabilities, quantum computing has the potential to revolutionize the reliability assessment of complex systems like large power grids. The "Quantum Twin" will be able provide high computational efficiency and flexibility for reliability assessment of complex systems.

One difference between classical reliability assessment using "Digital Twin" versus "Quantum Twin" is that Quantum Twin usually needs error correction due to the quantum decoherence and state fidelity. In addition, for large power systems, the number of components to consider state changes could be large, implying that a large number of qubits is needed to simulate large power systems. With the increase of qubits, the error rate also further increases, which may influence the reliability assessment results. Therefore, increasing the qubit number and reduce the error rate are important for reliability assessment of complex systems, which could be future work.